\documentstyle[prd,aps,preprint,tighten,epsfig]{revtex}

\begin{document}

\draft

\title{Implications of the KamLAND Measurement on
the Lepton Flavor Mixing Matrix and the Neutrino Mass Matrix}
\author{{\bf Wan-lei Guo} ~ and ~ {\bf Zhi-zhong Xing}}
\address{CCAST (World Laboratory), P.O. Box 8730,
Beijing 100039, China \\
and Institute of High Energy Physics, Chinese Academy of Sciences, \\
P.O. Box 918 (4), Beijing 100039, China
\footnote{Mailing address} \\
({\it Electronic address: guowl@mail.ihep.ac.cn;
xingzz@mail.ihep.ac.cn}) } \maketitle

\begin{abstract}
We explore some important implications of the KamLAND measurment 
on the lepton flavor mixing matrix $V$ and the neutrino mass matrix 
$M$. The model-independent constraints on nine matrix elements
of $V$ are obtained to a reasonable degree of accuracy. We find 
that nine two-zero textures of $M$ are compatible with current 
experimental data, but two of them are only marginally allowed.
Instructive predictions are given for the absolute neutrino
masses, Majorana phases of $CP$ violation, effective masses of
the tritium beta decay and neutrinoless double beta decay.
\end{abstract}


\section{Introduction}

The KamLAND experiment \cite{KM} turns to confirm the large-mixing-angle
(LMA) Mikheyev-Smirnov-Wolfenstein (MSW) solution \cite{MSW} to the
long-standing solar neutrino problem. In addition, the K2K long-baseline
experiment \cite{K2K} has unambiguously observed a reduction of $\nu_\mu$
flux and a distortion of the energy spectrum. These new measurements,
together with the compelling SNO evidence \cite{SNO} for the flavor
conversion of solar $\nu_e$ neutrinos and the Super-Kamiokande
evidecne \cite{SK} for the deficit of atmospheric $\nu_\mu$ neutrinos,
convinces us that the hypothesis of neutrino oscillations is actually
correct! We are then led to the conclusion that neutrinos are massive
and lepton flavors are mixed.

The mixing of lepton flavors means a mismatch between neutrino mass
eigenstates $(\nu_1, \nu_2, \nu_3)$ and neutrino flavor eigenstates
$(\nu_e, \nu_\mu, \nu_\tau)$ in the basis where the charged lepton
mass matrix is diagonal:
\begin{equation}
\left ( \matrix{
\nu_e \cr \nu_\mu \cr \nu_\tau \cr} \right )
\; =\; \left ( \matrix{
V_{e1} & V_{e2} & V_{e3} \cr
V_{\mu 1} & V_{\mu 2} & V_{\mu 3} \cr
V_{\tau 1} & V_{\tau 2} & V_{\tau 3} \cr} \right )
\left ( \matrix{
\nu_1 \cr \nu_2 \cr \nu_3 \cr} \right ) \; .
\end{equation}
The matrix elements $|V_{e1}|$, $|V_{e2}|$, $|V_{e3}|$ and $|V_{\mu 3}|$
can simply be related to the mixing factors of solar \cite{KM,SNO},
atmospheric \cite{SK} and CHOOZ reactor \cite{CHOOZ} neutrino
oscillations in the following way:
\begin{eqnarray}
\sin^2 2\theta_{\rm sun} & = & 4 |V_{e1}|^2 |V_{e2}|^2 \; ,
\nonumber \\
\sin^2 2\theta_{\rm atm} & = & 4 |V_{\mu 3}|^2
\left ( 1 - |V_{\mu 3}|^2 \right ) \; ,
\nonumber \\
\sin^2 2\theta_{\rm chz} & = & 4 |V_{e3}|^2
\left ( 1 - |V_{e3}|^2 \right ) \; .
\end{eqnarray}
Taking account of the unitarity of $V$, one may reversely express
$|V_{e1}|$, $|V_{e2}|$, $|V_{e3}|$, $|V_{\mu 3}|$ and $|V_{\tau 3}|$
in terms of $\theta_{\rm sun}$, $\theta_{\rm atm}$ and
$\theta_{\rm chz}$ \cite{Xing02a}:
\begin{eqnarray}
|V_{e1}| & = & \frac{1}{\sqrt 2} \sqrt{ \cos^2\theta_{\rm chz} +
\sqrt{\cos^4\theta_{\rm chz} - \sin^2 2\theta_{\rm sun}}} \;\; ,
\nonumber \\
|V_{e2}| & = & \frac{1}{\sqrt 2} \sqrt{ \cos^2\theta_{\rm chz} -
\sqrt{\cos^4\theta_{\rm chz} - \sin^2 2\theta_{\rm sun}}} \;\; ,
\nonumber \\
|V_{e3}| & = & \sin\theta_{\rm chz} \; ,
\nonumber \\
|V_{\mu 3}| & = & \sin\theta_{\rm atm} \; ,
\nonumber \\
|V_{\tau 3}| & = & \sqrt{\cos^2\theta_{\rm chz} -
\sin^2\theta_{\rm atm}} \; .
\end{eqnarray}
Current experimental information on $\theta_{\rm sun}$,
$\theta_{\rm atm}$ and $\theta_{\rm chz}$ allows us to get very
instructive constraints on the lepton flavor mixing matrix $V$.
One purpose of this paper is therefore to examine how accurately we
can recast $V$ from the present KamLAND, K2K, SNO, Super-Kamiokande
and CHOOZ measurements.

Another purpose of this paper is to confront two-zero textures of
the neutrino mass matrix with the new KamLAND data, so as to single
out the most favorable texture(s) in phenomenology. In the flavor
basis chosen above, the Majorana neutrino mass matrix can be
written as
\begin{equation}
M \; =\; V \left ( \matrix{
m_1 & 0 & 0 \cr
0 & m_2 & 0 \cr
0 & 0 & m_3 \cr} \right ) V^{\rm T} \; ,
\end{equation}
where $m_i$ (for $i=1,2,3$) are physical masses of three neutrinos.
In the assumption of the LMA solution for solar neutrino
oscillations, a classification of $M$ with two vanishing entries has
been done in an analytically approximate way \cite{Glashow,Xing02b}. 
Our present analysis is different from
the previous ones in two important aspects: (1) we carry out a careful
numerical analysis of every two-zero pattern of the neutrino mass
matrix $M$ to pin down its complete parameter space, because simple
analytical approximations are sometimes unable to reveal the whole
regions of relevant parameters allowed by new experimental data;
(2) we present the quantitative predictions for allowed ranges of
the absolute neutrino masses, the Majorana phases of $CP$ violation,
and the effective masses of the tritium beta decay
($\langle m\rangle_e$) and neutrinoless double beta
decay ($\langle m\rangle_{ee}$).

The remaining part of this paper is organized as follows. With the 
help of current data, we derive
the model-independent constraints on nine elements of the lepton
flavor mixing matrix in section II. Section III is devoted to a
detailed analysis of the parameter space for every two-zero texture
of the neutrino mass matrix. We obtain instructive predictions for the
neutrino mass spectrum and Majorana phases of $CP$ violation as well
as $\langle m\rangle_e$ and $\langle m\rangle_{ee}$ in
section IV. Finally a brief summary is given in section V.

\section{Constraints on the lepton flavor mixing matrix}

As already shown in Eq. (3), five matrix elements of $V$ can be
determined or constrained from current experimental data. The other
four matrix elements ($|V_{\mu 1}|$, $|V_{\mu 2}|$, $|V_{\tau 1}|$
and $|V_{\tau 2}|$) are entirely unrestricted, however, unless one
of them or the rephasing invariant of $CP$ violation of $V$ (defined
as $J$ \cite{J}) is measured. A realistic way to get rough but useful
constraints on those four unknown elements is to allow the Dirac
phase of $CP$ violation in $V$ to vary between 0 and
$\pi$ \cite{Tanimoto}, such that one can find out the maximal and
minimal magnitudes of each matrix element. To see this point more
clearly, we adopt the standard parametrization $V = U P$ \cite{FX01},
where
\begin{equation}
U \; = \; \left ( \matrix{
c_x c_z & s_x c_z & s_z \cr
- c_x s_y s_z - s_x c_y e^{-i\delta} &
- s_x s_y s_z + c_x c_y e^{-i\delta} &
s_y c_z \cr
- c_x c_y s_z + s_x s_y e^{-i\delta} &
- s_x c_y s_z - c_x s_y e^{-i\delta} &
c_y c_z \cr } \right ) \;
\end{equation}
with $s_x \equiv \sin\theta_x$, $c_x \equiv \cos\theta_x$, and
so on; and
\begin{equation}
P \; = \; \left ( \matrix{
e^{i\rho} & 0 & 0 \cr
0 & e^{i\sigma} & 0 \cr
0 & 0 & 1 \cr} \right ) \; .
\end{equation}
The advantage of this representation is that the neutrinoless
double beta decay is associated with the Majorana phases $\rho$
and $\sigma$, while $CP$ violation in normal neutrino oscillations
depends separately on the Dirac phase $\delta$. Note that three
mixing angles $(\theta_x, \theta_y, \theta_z)$, which are all
arranged to lie in the first quadrant, can be written as
\begin{eqnarray}
\tan\theta_x & = & \frac{|V_{e2}|}{|V_{e1}|} \; ,
\nonumber \\
\tan\theta_y & = & \frac{|V_{\mu 3}|}{|V_{\tau 3}|} \; ,
\nonumber \\
\sin\theta_z & = & |V_{e3}| \; .
\end{eqnarray}
It is then straightforward to obtain
\begin{eqnarray}
|V_{\mu 1}| & = & \frac{\left | |V_{e2}| |V_{\tau 3}| +
|V_{e1}| |V_{e3}| |V_{\mu 3}| ~ e^{i\delta} \right |}
{1 - |V_{e3}|^2} \; ,
\nonumber \\
|V_{\mu 2}| & = & \frac{\left | |V_{e1}| |V_{\tau 3}| -
|V_{e2}| |V_{e3}| |V_{\mu 3}| ~ e^{i\delta} \right |}
{1 - |V_{e3}|^2} \; ,
\nonumber \\
|V_{\tau 1}| & = & \frac{\left | |V_{e2}| |V_{\mu 3}| -
|V_{e1}| |V_{e3}| |V_{\tau 3}| ~ e^{i\delta} \right |}
{1 - |V_{e3}|^2} \; ,
\nonumber \\
|V_{\tau 2}| & = & \frac{\left | |V_{e1}| |V_{\mu 3}| +
|V_{e2}| |V_{e3}| |V_{\tau 3}| ~ e^{i\delta} \right |}
{1 - |V_{e3}|^2} \; .
\end{eqnarray}
Varying the Dirac phase $\delta$ from 0 to $\pi$, we are led to
the {\it most generous} ranges of $|V_{\mu 1}|$, $|V_{\mu 2}|$,
$|V_{\tau 1}|$ and $|V_{\tau 2}|$:
\begin{eqnarray}
\frac{|V_{e2}| |V_{\tau 3}| - |V_{e1}| |V_{e3}| |V_{\mu 3}|}
{1 - |V_{e3}|^2} & \leq & |V_{\mu 1}| \; \leq \;
\frac{|V_{e2}| |V_{\tau 3}| + |V_{e1}| |V_{e3}| |V_{\mu 3}|}
{1 - |V_{e3}|^2} \; ,
\nonumber \\
\frac{|V_{e1}| |V_{\tau 3}| - |V_{e2}| |V_{e3}| |V_{\mu 3}|}
{1 - |V_{e3}|^2} & \leq & |V_{\mu 2}| \; \leq \;
\frac{|V_{e1}| |V_{\tau 3}| + |V_{e2}| |V_{e3}| |V_{\mu 3}|}
{1 - |V_{e3}|^2} \; ,
\nonumber \\
\frac{|V_{e2}| |V_{\mu 3}| - |V_{e1}| |V_{e3}| |V_{\tau 3}|}
{1 - |V_{e3}|^2} & \leq & |V_{\tau 1}| \; \leq \;
\frac{|V_{e2}| |V_{\mu 3}| + |V_{e1}| |V_{e3}| |V_{\tau 3}|}
{1 - |V_{e3}|^2} \; ,
\nonumber \\
\frac{|V_{e1}| |V_{\mu 3}| - |V_{e2}| |V_{e3}| |V_{\tau 3}|}
{1 - |V_{e3}|^2} & \leq & |V_{\tau 2}| \; \leq \;
\frac{|V_{e1}| |V_{\mu 3}| + |V_{e2}| |V_{e3}| |V_{\tau 3}|}
{1 - |V_{e3}|^2} \; .
\end{eqnarray}
Note that the lower and upper bounds of each matrix element turn to
coincide with each other in the limit $|V_{e3}| \rightarrow 0$. Because of
the smallness of $|V_{e3}|$, the ranges obtained in Eq. (9) should be quite
restrictive. Hence it makes sense to recast the lepton flavor mixing matrix
even in the absence of any experimental information on $CP$ violation.

In view of the present experimental data from KamLAND \cite{KM},
K2K \cite{K2K}, SNO \cite{SNO}, Super-Kamiokande \cite{SK} and
CHOOZ \cite{CHOOZ}, we have $0.25 \leq \sin^2\theta_{\rm sun} \leq
0.40$ \cite{Fogli}, $0.92 < \sin^2 2\theta_{\rm atm} \leq 1.0$, and $0 \leq
\sin^2 2\theta_{\rm chz} < 0.1$ at the $90\%$ confidence level.
Namely,
\begin{eqnarray}
30.0^\circ & \leq & \theta_{\rm sun} \; \leq \; 39.2^\circ \; ,
\nonumber \\
36.8^\circ & < & \theta_{\rm atm} \; < \; 53.2^\circ \; ,
\nonumber \\
0^\circ & \leq & \theta_{\rm chz} \; < \; 9.2^\circ \; .
\end{eqnarray}
Using these inputs, we calculate the numerical ranges of
$|V_{e1}|$, $|V_{e2}|$, $|V_{e3}|$, $|V_{\mu 3}|$ and $|V_{\tau 3}|$
from Eq. (3). Then the allowed ranges of $|V_{\mu 1}|$, $|V_{\mu 2}|$,
$|V_{\tau 1}|$ and $|V_{\tau 2}|$ can be found with the help of
Eq. (9). Our numerical results are summarized as
\begin{equation}
|V| \; =\; \left ( \matrix{ 0.70-0.87 & 0.50-0.69 & <0.16 \cr
0.20-0.61 & 0.34-0.73 & 0.60-0.80 \cr 0.21-0.63 & 0.36-0.74 &
0.58-0.80 \cr} \right ) \; .
\end{equation}
This result is certainly more restrictive than that obtained in
Ref. \cite{Tanimoto} before the KamLAND measurement.

Note that the rephasing invariant of $CP$ violation reads as follows:
\begin{eqnarray}
J & = & \frac{|V_{e1}| |V_{e2}| |V_{e3}| |V_{\mu 3}| |V_{\tau 3}|}
{1 - |V_{e3}|^2} \sin\delta
\nonumber \\
& = & \frac{\sqrt{\sin^2 2\theta_{\rm sun}
\left ( \sin^2 2\theta_{\rm atm}
+ 4 \sin^2\theta_{\rm atm} \sin^2\theta_{\rm chz} \right )}}
{4 \cos^2\theta_{\rm chz}} \sin\theta_{\rm chz} \sin\delta \; .
\end{eqnarray}
The term proportional to $4\sin^2\theta_{\rm atm}
\sin^2\theta_{\rm chz}$ in $J$, which may correct the leading term
up to $5\%$ (for $\theta_{\rm atm} = 45^\circ$ and $\theta_{\rm
chz} = 9^\circ$), were not taken into account in Ref.
\cite{Tanimoto}. By use of Eq. (10), we find $J \leq
0.039\sin\delta$. This result implies that the magnitude of $J$
can maximally be 0.039, leading probably to observable
$CP$-violating effects in long-baseline neutrino oscillations.

\section{Two-zero textures of the neutrino mass matrix}

The symmetric neutrino mass matrix $M$ totally has six independent
complex entries. If two of them vanish, i.e.,
$M_{ab} = M_{pq} =0$, we obtain two constraint equations:
\begin{eqnarray}
m_1 U_{a1} U_{b1} e^{2i\rho} + m_2 U_{a2} U_{b2} e^{2i\sigma}
+ m_3 U_{a3} U_{b3} & = & 0 \; ,
\nonumber \\
m_1 U_{p1} U_{q1} e^{2i\rho} + m_2 U_{p2} U_{q2} e^{2i\sigma}
+ m_3 U_{p3} U_{q3} & = & 0 \; ,
\end{eqnarray}
where $a$, $b$, $p$ and $q$ run over $e$, $\mu$ and $\tau$, but
$(p,q) \neq (a,b)$. Solving Eq. (12), we arrive at \cite{Xing02b}
\begin{eqnarray}
\frac{m_1}{m_3} e^{2i\rho} & = &
\frac{U_{a3} U_{b3} U_{p2} U_{q2} - U_{a2} U_{b2} U_{p3}
U_{q3}}{U_{a2} U_{b2} U_{p1} U_{q1} - U_{a1} U_{b1}U_{p2} U_{q2}} \; ,
\nonumber \\
\frac{m_2}{m_3} e^{2i\sigma} & = &
\frac{U_{a1} U_{b1} U_{p3} U_{q3} - U_{a3} U_{b3} U_{p1}
U_{q1}}{U_{a2} U_{b2} U_{p1} U_{q1} - U_{a1} U_{b1}U_{p2} U_{q2}} \; .
\end{eqnarray}
This result implies that two neutrino mass ratios
$(m_1/m_3, m_2/m_3)$ and two Majorana-type $CP$-violating phases
$(\rho, \sigma)$ can fully be determined in terms of three mixing angles
$(\theta_x, \theta_y, \theta_z)$ and the Dirac-type $CP$-violating phase
$(\delta)$. Thus one may examine whether a two-zero texture of $M$ is
empirically acceptable or not by comparing its prediction for the
ratio of two neutrino mass-squared differences with the result extracted
from current experimental data on solar and atmospheric neutrino
oscillations:
\begin{equation}
R_\nu \; \equiv \; \frac{\left |m^2_2 - m^2_1 \right |}
{\left |m^2_3 - m^2_2 \right |} \; \approx \;
\frac{\Delta m^2_{\rm sun}}{\Delta m^2_{\rm atm}} \; .
\end{equation}
Considering the LMA MSW solution confirmed by the KamLAND measurement, 
we have $5.9 \times 10^{-5} ~ {\rm eV^2} \leq \Delta m^2_{\rm sun} 
\leq 8.8 \times 10^{-5} ~ {\rm eV^2}$ \cite{Fogli,Barger}
at the $90\%$ confidence level. In addition, we have 
$1.6 \times 10^{-3} ~ {\rm eV^2} \leq
\Delta m^2_{\rm atm} \leq 3.9 \times 10^{-3} ~ {\rm eV^2}$
\cite{ATM} at the $90\%$ confidence level. Thus we arrive at 
$1.5 \times 10^{-2} \leq R_\nu \leq 5.5 \times 10^{-2}$. The allowed
ranges of three mixing angles $\theta_x \approx \theta_{\rm sun}$,
$\theta_y \approx \theta_{\rm atm}$ and $\theta_z \approx
\theta_{\rm chz}$ have been given in Eq. (10). There is no
experimental constraint on the $CP$-violating phase $\delta$.
Hence we simply take $\delta$ from $0^\circ$ to $360^\circ$ in our
numerical calculations.

There are totally fifteen distinct
topologies for the structure of $M$ with two independent vanishing
entries, as shown in Tables 1 and 2. We work out the explicit
expressions of $(m_1/m_3)e^{2i\rho}$ and $(m_2/m_3)e^{2i\sigma}$ for
each pattern of $M$ by use of Eq. (14), and list the results in
the same tables \cite{Guo}. With the input values of $\theta_x$,
$\theta_y$, $\theta_z$ and $\delta$ mentioned above, we calculate
the ratio $R_\nu$ and examine whether it is in the range allowed
by current data. This criterion has been used in 
Refs. \cite{Glashow,Xing02b} to pick the phenomenologically favored 
patterns of $M$ in the LMA case.

Nine of the fifteen two-zero textures of $M$ listed in Table 1 are
found to be in accord with the LMA solution as well as the
atmospheric neutrino data. They can be classified into four
categories: A (with $\rm A_1$ and $\rm A_2$), B (with $\rm B_1$,
$\rm B_2$, $\rm B_3$ and $\rm B_4$), C and D (with $\rm D_1$ and
$\rm D_2$). The point of this classification is that the textures
of $M$ in each category result in similar physical consequences,
which are almost indistinguishable in practice. The other six
patterns of $M$ (categories E and F) listed in Table 2 cannot
coincide with current experimental data. In particular, the exact
neutrino mass degeneracy ($m_1 = m_2 = m_3$) is predicted from
three textures of $M$ belonging to category F.

Now let us focus on patterns $\rm A_1$, $\rm B_1$, $\rm C$ and
$\rm D_1$ as four typical examples for numerical illustration. Our
results for $\sin^2 2\theta_{\rm chz}$ versus $\delta$ and
$\theta_y$ versus $\theta_x$ are shown Figs. 1 -- 4. Some comments
are in order.

(1) For pattern $\rm A_1$, arbitrary values of $\delta$ are
allowed if $\sin^2 2\theta_{\rm chz}$ is large enough ($\geq
0.014$). The mixing angles $\theta_x$ and $\theta_y$ may take any
values in the ranges allowed by current data. Therefore we
conclude that pattern $\rm A_1$ is favored in phenomenology with
little fine-tuning. A similar conclusion can be drawn for pattern
$\rm A_2$.

(2) For pattern $\rm B_1$, $\delta$ is essentially unconstrained
if $\sin^2 2\theta_{\rm chz}$ is extremely close to zero; and only
$\delta$ around $90^\circ$ or $ 270^\circ$ is acceptable if
$\sin^2 2\theta_{\rm chz}$ deviates somehow from zero. Except
$\theta_y \neq 45^\circ$, there is no further constraint on the
parameter space of $(\theta_x, \theta_y)$. We conclude that
pattern $\rm B_1$ with maximal $CP$ violation (i.e., $\sin\delta
\approx \pm 1$) is phenomenologically favored. So are patterns
$\rm B_2$, $\rm B_3$ and $\rm B_4$.

(3) For pattern $\rm C$, $\delta = 90^\circ$ or
$\delta = 270^\circ$ is forbidden. Furthermore,
$\theta_y = 45^\circ$ is forbidden. We see that the allowed
parameter space of $(\delta, \theta_{\rm chz})$ and that of
$(\theta_x, \theta_y)$ are rather large. Hence pattern C is
also favored in phenomenology.

(4) For pattern $\rm D_1$, $\delta$ is restricted to be around
$0^\circ$ or $360^\circ$. In particular, the region $90^\circ \leq
\delta \leq 270^\circ$ is entirely excluded. $\sin^2 2\theta_{\rm
chz} > 0.084$ holds for the allowed range of $\delta$. Different
from patterns $\rm A_1$, $\rm B_1$ and $\rm C$, pattern $\rm D_1$
requires relatively strong correlation between $\theta_x$ and
$\theta_y$ (e.g., small values of $\theta_y$ are associated with
large values of $\theta_x$ in the allowed parameter space). In
this sense, we argue that pattern $\rm D_1$ is less natural in
phenomenology, although it has not been ruled out by current
experimental data. A similar argument can be made for pattern 
$\rm D_2$.

It is worth remarking that patterns $\rm D_1$ and $\rm D_2$ were
not included into the phenomenologically allowed patterns of $M$
in the previous classification \cite{Glashow,Xing02b}, where only 
analytical approximations were made. Our numerical analysis shows
that these two patterns are marginally allowed by current data.
Here we have also explored some interesting details of the parameter 
space for every favorable texture, which could not be seen from
simple analytical approximations.

\section{Numerical predictions and further discussions}

A two-zero texture of $M$ has a number of
interesting predictions, in particular, for the absolute neutrino
masses and the Majorana phases of $CP$ violation \cite{Xing02b}.
With the help of Eq. (14), one may calculate the mass ratios
$m_1/m_3$ and $m_2/m_3$ as well as the Majorana phases $\rho$ and
$\sigma$. The absolute neutrino mass $m_3$ can be determined from
\begin{equation}
m_3 \; =\; \frac{1}{\sqrt{\displaystyle \left | 1 -
\left (\frac{m_2}{m_3} \right )^2 \right |}}
~ \sqrt{\Delta m^2_{\rm atm}} \;\; .
\end{equation}
Therefore a full determination of the mass spectrum of three
neutrinos is actually possible. Then we may obtain definite
predictions for the effective mass of the tritium beta decay,
\begin{equation}
\langle m\rangle_e \; =\; m_1 c^2_x c^2_z + m_2 s^2_x c^2_z +
m_3 s^2_z \; ;
\end{equation}
and that of the neutrinoless double beta decay,
\begin{equation}
\langle m\rangle_{ee} \; =\; \left | m_1 c^2_x c^2_z e^{2i\rho}
+ m_2 s^2_x c^2_z e^{2i\sigma} + m_3 s^2_z \right | \; .
\end{equation}
It is clear that the Dirac phase $\delta$ has no contribution to
$\langle m\rangle_{ee}$. Note that $CP$- and $T$-violating
asymmetries in normal neutrino oscillations are controlled by
$\delta$ or the rephasing-invariant parameter
$J = s_x c_x s_y c_y s_z c^2_z\sin\delta$. Whether
$\langle m\rangle_e$ and $\langle m\rangle_{ee}$ can be measured
remains an open question. The present experimental upper bounds
are $\langle m\rangle_e < 2.2 ~ {\rm eV}$ \cite{PDG} and $\langle
m\rangle_{ee} < 0.35 ~ {\rm eV}$ \cite{HM} at the $90\%$
confidence level. The proposed KATRIN experiment is possible to 
reach the sensitivity 
$\langle m\rangle_e \sim 0.3 ~ {\rm eV}$ \cite{KATRIN}, and a 
number of next-generation experiments for the
neutrinoless double beta decay \cite{DB} is possible to probe
$\langle m\rangle_{ee}$ at the level of 10 meV to 50 meV.

We perform a numerical calculation of $m_2/m_3$ versus $m_1/m_3$,
$\sigma$ versus $\rho$, $\langle m\rangle_{ee}$ versus $\langle
m\rangle_e$, and $J$ versus $m_3$ for patterns $\rm A_1$, 
$\rm B_1$, $\rm C$ and $\rm D_1$. The results are shown in 
Figs. 1 -- 4. Some discussions are in order.

(1) For pattern $\rm A_1$, $\rho \approx \delta/2$ or $\rho
\approx \delta/2 - 180^\circ$ and $\sigma \approx \rho \pm
90^\circ$ hold in most cases. Two neutrino mass ratios lie in the
ranges $0.033 \leq m_1/m_3 \leq 0.19$ and $0.13 \leq m_2/m_3
\leq 0.28$, and the absolute value of $m_3$ is in the range
$0.04 ~ {\rm eV} \leq m_3 \leq 0.065 ~ {\rm eV}$. As $\langle
m\rangle_{ee} = 0$ is a direct consequence of texture $\rm A_1$,
we calculate the sum of three neutrino masses $\sum m_i$ instead
of $\langle m\rangle_{ee}$. The result is $0.047 ~ {\rm eV} \leq
\sum m_i \leq 0.093 ~ {\rm eV}$, in contrast with $0.003 ~ {\rm
eV} \leq \langle m\rangle_e \leq 0.014 ~ {\rm eV}$. The rephasing
invariant of $CP$ violation $J$ is found to lie in the range
$-0.037 \leq J \leq 0.038$. Similar predictions are expected for
pattern $\rm A_2$.

(2) For pattern $\rm B_1$, $\rho \approx \sigma \approx \delta -
90^\circ$ or $\rho \approx \sigma \approx \delta - 270^\circ$
holds in most cases. Two neutrino mass ratios $m_1/m_3$ may lie
either in the range $0.53 \leq m_1/m_3  \leq 0.99$ or in the
range $1.01 \leq m_1/m_3 \leq 1.88$, and $m_2/m_3$ may lie
either in the range $0.53 \leq m_2/m_3  \leq 0.99$ or in the
range $1.01 \leq m_2/m_3 \leq 1.88$.  The value of $m_3$ is
found to be in the range $0.026 ~ {\rm eV} \leq m_3 \leq 0.35 ~
{\rm eV}$. Furthermore, we arrive at $0.027 ~ {\rm eV} \leq
\langle m\rangle_e \approx \langle m\rangle_{ee} \leq 0.35 ~ {\rm
eV}$ as well as $-0.037 \leq J \leq 0.037$. Similar results can be
obtained for patterns $\rm B_2$, $\rm B_3$ and $\rm B_4$.

(3) For pattern $\rm C$, $\sigma \approx \rho$ when $\theta_y$
approaches $45^\circ$; and there is no clear correlation between
$\rho$ and $\sigma$ for other values of $\theta_y$. Two neutrino
mass ratios $m_1/m_3$ may lie either in the range $0.95 \leq
m_1/m_3  \leq 0.99$ or in the range $1.01 \leq m_1/m_3 \leq
5.4$, and $m_2/m_3$ may lie either in the range $0.95 \leq
m_2/m_3 \leq 0.99$ or in the range $1.01 \leq m_2/m_3 \leq
5.3$. The value of $m_3$ is found to lie in the range $0.009 ~
{\rm eV} \leq m_3 \leq 0.35 ~ {\rm eV}$. It is remarkable that
$\langle m\rangle_{ee} \approx m_3$ holds to a good degree of
accuracy in the allowed space of those input parameters. We also
obtain $0.04 ~ {\rm eV} \leq \langle m\rangle_e \leq 0.35 ~ {\rm
eV}$ and $-0.037 \leq J \leq 0.037$.

(4) For pattern $\rm D_1$, $\rho \approx \delta - 90^\circ$ or
 $\rho \approx \delta - 270^\circ$ and
$\sigma \approx \rho \pm 90^\circ$ hold. Two neutrino mass ratios
lie in the ranges $7.5 \leq m_1/m_3 \leq 8.8$ and $7.35 \leq
m_2/m_3 \leq 8.6$, and the absolute value of $m_3$ is in the
range $0.005 ~ {\rm eV} \leq m_3 \leq 0.008 ~ {\rm eV}$. As for
the tritium beta decay and neutrinoless double beta decay, we
obtain $0.04 ~ {\rm eV} \leq \langle m\rangle_e \leq 0.062 ~ {\rm
eV}$ and $0.008 ~ {\rm eV} \leq \langle m\rangle_{ee} \leq 0.014 ~
{\rm eV}$. The range of $J$ is found to be $-0.014 \leq J \leq
0.011$. Similar predictions can straightforwardly be made for
pattern $\rm D_2$.

We see that there is no hope to measure both $\langle m\rangle_e$
and $\langle m\rangle_{ee}$, if the neutrino mass matrix $M$ takes
pattern $\rm A_1$ or $\rm A_2$. As for categories B and C of $M$,
the upper limit of $\langle m\rangle_e$ is close to the
sensitivity of the KATRIN experiment ($\sim 0.3 ~ {\rm eV}$
\cite{KATRIN}), and that of $\langle m\rangle_{ee}$ is just below
the current experimental bound \cite{HM}.

\section{Summary}

In summary, we have discussed some implications of the KamLAND
measurment on the lepton flavor mixing matrix $V$ and the neutrino
mass matrix $M$. The model-independent constraints on nine elements
of $V$ have been obtained up to a reasonable degree of accuracy.
Nine two-zero textures of $M$ are found to be compatible with current
experimental data, but two of them are only marginally allowed.
Instructive consequences of these phenomenologically favored textures
of $M$ on the absolute neutrino masses, Majorana phases of 
$CP$ violation, $\langle m\rangle_e$ and $\langle m\rangle_{ee}$ are 
numerically explored. Our results will be very useful for model 
building \cite{Review}, in order to understand why neutrino masses 
are so tiny and why two of the lepton flavor mixing angles are so large.

Finally it is worth remarking that a specific texture of lepton mass
matrices may not be preserved to all orders or at any energy scales in
the unspecified interactions from which lepton masses are generated.
Nevertheless, those phenomenologically favored textures at low energy
scales, no matter whether they are of the two-zero form or other
forms, are possible to provide enlightening hints at the underlying
dynamics of lepton mass generation at high energy scales.

\vspace{0.5cm}

We would like to thank Y.F. Wang for stimulating discussions about the
KamLAND measurement. One of us (Z.Z.X.) is also grateful to IPPP in
University of Durham, where the paper was finalized, for its warm
hospitality. This work was supported in part by National Natural
Science Foundation of China.

\newpage

\begin{table}
\caption{Nine patterns of the neutrino mass matrix $M$ with two independent
vanishing entries, which are {\it compatible} with the LMA solution and other
empirical hypotheses. The analytical results for two ratios of three neutrino
mass eigenvalues $(m_1/m_3)e^{2i\rho}$ and $(m_2/m_3)e^{2i\sigma}$ are
given in terms of four flavor mixing parameters $\theta_x$, $\theta_y$,
$\theta_z$ and $\delta$.}
\begin{center}
\begin{tabular}{ll} 
Pattern of $M$ & Results of $(m_1/m_3)e^{2i\rho}$ and 
$(m_2/m_3)e^{2i\sigma}$  \\ \hline \\
$\rm A_1:
\left ( \matrix{
{\bf 0} & {\bf 0} & \times \cr
{\bf 0} & \times & \times \cr
\times & \times & \times \cr} \right )$
&
$\matrix{
\frac{m_1}{m_3} e^{2i\rho} =
+ \frac{s_z}{c^2_z} \left ( \frac{s_x s_y}{c_x c_y} ~ e^{i\delta}
- s_z \right ) \cr
\frac{m_2}{m_3} e^{2i\sigma} =
- \frac{s_z}{c^2_z} \left ( \frac{c_x s_y}{s_x c_y} ~ e^{i\delta}
+ s_z \right ) }$
\\ \\
$\rm A_2:
\left ( \matrix{
{\bf 0} & \times & {\bf 0} \cr
\times & \times & \times \cr
{\bf 0} & \times & \times \cr} \right )$
&
$\matrix{
\frac{m_1}{m_3} e^{2i\rho} =
- \frac{s_z}{c^2_z} \left ( \frac{s_x c_y}{c_x s_y} ~ e^{i\delta}
+ s_z \right ) \cr
\frac{m_2}{m_3} e^{2i\sigma} =
+ \frac{s_z}{c^2_z} \left ( \frac{c_x c_y}{s_x s_y} ~ e^{i\delta}
- s_z \right ) }$
\\ \\
$\rm B_1:
\left ( \matrix{
\times & \times & {\bf 0} \cr
\times & {\bf 0} & \times \cr
{\bf 0} & \times & \times \cr} \right )$
&
$\matrix{
\frac{m_1}{m_3} e^{2i\rho} =
\frac{s_x c_x s_y \left (2 c^2_y s^2_z - s^2_y c^2_z \right )
- c_y s_z \left ( s^2_x s^2_y e^{+i\delta} + c^2_x c^2_y e^{-i\delta} \right )}
{s_x c_x s_y c^2_y +
\left ( s^2_x - c^2_x \right ) c^3_y s_z e^{i\delta} +
s_x c_x s_y s^2_z \left ( 1 + c^2_y \right ) e^{2i\delta}} ~ e^{2i\delta} \cr
\frac{m_2}{m_3} e^{2i\sigma} =
\frac{s_x c_x s_y \left (2 c^2_y s^2_z - s^2_y c^2_z \right )
+ c_y s_z \left ( c^2_x s^2_y e^{+i\delta} + s^2_x c^2_y e^{-i\delta} \right )}
{s_x c_x s_y c^2_y +
\left ( s^2_x - c^2_x \right ) c^3_y s_z e^{i\delta} +
s_x c_x s_y s^2_z \left ( 1 + c^2_y \right ) e^{2i\delta}} ~ e^{2i\delta} }$
\\ \\
$\rm B_2:
\left ( \matrix{
\times & {\bf 0} & \times \cr
{\bf 0} & \times & \times \cr
\times & \times & {\bf 0} \cr} \right )$
&
$\matrix{
\frac{m_1}{m_3} e^{2i\rho} =
\frac{s_x c_x c_y \left (2 s^2_y s^2_z - c^2_y c^2_z \right )
+ s_y s_z \left ( s^2_x c^2_y e^{+i\delta} + c^2_x s^2_y e^{-i\delta} \right )}
{s_x c_x s^2_y c_y -
\left ( s^2_x - c^2_x \right ) s^3_y s_z e^{i\delta} +
s_x c_x c_y s^2_z \left ( 1 + s^2_y \right ) e^{2i\delta}} ~ e^{2i\delta} \cr
\frac{m_2}{m_3} e^{2i\sigma} =
\frac{s_x c_x c_y \left (2 s^2_y s^2_z - c^2_y c^2_z \right )
- s_y s_z \left ( c^2_x c^2_y e^{+i\delta} + s^2_x s^2_y e^{-i\delta} \right )}
{s_x c_x s^2_y c_y -
\left ( s^2_x - c^2_x \right ) s^3_y s_z e^{i\delta} +
s_x c_x c_y s^2_z \left ( 1 + s^2_y \right ) e^{2i\delta}} ~ e^{2i\delta} }$
\\ \\
$\rm B_3:
\left ( \matrix{
\times & {\bf 0} & \times \cr
{\bf 0} & {\bf 0} & \times \cr
\times & \times & \times \cr} \right )$
&
$\matrix{
\frac{m_1}{m_3} e^{2i\rho} =
- \frac{s_y}{c_y} \cdot \frac{s_x s_y - c_x c_y s_z e^{-i\delta}}
{s_x c_y + c_x s_y s_z e^{+i\delta}} ~ e^{2i\delta} \cr
\frac{m_2}{m_3} e^{2i\sigma} =
- \frac{s_y}{c_y} \cdot \frac{c_x s_y + s_x c_y s_z e^{-i\delta}}
{c_x c_y - s_x s_y s_z e^{+i\delta}} ~ e^{2i\delta} }$
\\ \\
$\rm B_4:
\left ( \matrix{
\times & \times & {\bf 0} \cr
\times & \times & \times \cr
{\bf 0} & \times & {\bf 0} \cr} \right )$
&
$\matrix{
\frac{m_1}{m_3} e^{2i\rho} =
- \frac{c_y}{s_y} \cdot \frac{s_x c_y + c_x s_y s_z e^{-i\delta}}
{s_x s_y - c_x c_y s_z e^{+i\delta}} ~ e^{2i\delta} \cr
\frac{m_2}{m_3} e^{2i\sigma} =
- \frac{c_y}{s_y} \cdot \frac{c_x c_y - s_x s_y s_z e^{-i\delta}}
{c_x s_y + s_x c_y s_z e^{+i\delta}} ~ e^{2i\delta} }$
\\ \\
$\rm C: ~
\left ( \matrix{
\times & \times & \times \cr
\times & {\bf 0} & \times \cr
\times & \times & {\bf 0} \cr} \right )$
&
$\matrix{
\frac{m_1}{m_3} e^{2i\rho} =
- \frac{c_x c^2_z}{s_z} \cdot \frac{ c_x \left ( s^2_y - c^2_y \right )
+ 2 s_x s_y c_y s_z e^{i\delta}}
{2 s_x c_x s_y c_y - \left ( s^2_x - c^2_x \right )
\left ( s^2_y - c^2_y \right ) s_z e^{i\delta} + 2 s_x c_x s_y c_y s^2_z
e^{2i\delta}} ~ e^{i\delta} \cr
\frac{m_2}{m_3} e^{2i\sigma} =
+ \frac{s_x c^2_z}{s_z} \cdot \frac{ s_x \left ( s^2_y - c^2_y \right )
- 2 c_x s_y c_y s_z e^{i\delta}}
{2 s_x c_x s_y c_y - \left ( s^2_x - c^2_x \right )
\left ( s^2_y - c^2_y \right ) s_z e^{i\delta} + 2 s_x c_x s_y c_y s^2_z
e^{2i\delta}} ~ e^{i\delta} }$ 
\\ \\
$\rm D_1:
\left ( \matrix{
\times & \times & \times \cr
\times & {\bf 0} & {\bf 0} \cr
\times & {\bf 0} & \times \cr} \right )$
&
$\matrix{
\frac{m_1}{m_3} e^{2i\rho} =
- \frac{c^2_z}{s_z} \cdot \frac{c_x s_y}{s_x c_y +
c_x s_y s_z e^{i\delta}} ~ e^{i\delta} \cr
\frac{m_2}{m_3} e^{2i\sigma} =
+ \frac{c^2_z}{s_z} \cdot \frac{s_x s_y}{c_x c_y -
s_x s_y s_z e^{i\delta}} ~ e^{i\delta} }$
\\ \\
$\rm D_2:
\left ( \matrix{
\times & \times & \times \cr
\times & \times & {\bf 0} \cr
\times & {\bf 0} & {\bf 0} \cr} \right )$
&
$\matrix{
\frac{m_1}{m_3} e^{2i\rho} =
+ \frac{c^2_z}{s_z} \cdot \frac{c_x c_y}{s_x s_y -
c_x c_y s_z e^{i\delta}} ~ e^{i\delta} \cr
\frac{m_2}{m_3} e^{2i\sigma} =
- \frac{c^2_z}{s_z} \cdot \frac{s_x c_y}{c_x s_y +
s_x c_y s_z e^{i\delta}} ~ e^{i\delta} }$ \\ \\
\end{tabular}
\end{center}
\end{table}

\begin{table}
\caption{Six patterns of the neutrino mass matrix $M$ with two independent
vanishing entries, which are {\it incompatible} with the LMA solution and other
empirical hypotheses. The analytical results for two ratios of three neutrino
mass eigenvalues $(m_1/m_3)e^{2i\rho}$ and $(m_2/m_3)e^{2i\sigma}$ are
given in terms of four flavor mixing parameters $\theta_x$, $\theta_y$,
$\theta_z$ and $\delta$.}
\begin{center}
\begin{tabular}{ll} 
Pattern of $M$ & Results of $(m_1/m_3)e^{2i\rho}$ and 
$(m_2/m_3)e^{2i\sigma}$ \\ \hline \\
$\rm E_1:
\left ( \matrix{
{\bf 0} & \times & \times \cr
\times  & {\bf 0} & \times \cr
\times & \times & \times \cr} \right )$
&
$\matrix{
\frac{m_1}{m_3} e^{2i\rho} =
- \frac{1}{c_y c^2_z} \cdot \frac{s^2_x s^2_y \left (c^2_z - s^2_z \right )
- c_x c_y s^2_z \left (c_x c_y - 2 s_x s_y s_z e^{i\delta} \right ) e^{-2i\delta}}
{\left (s^2_x - c^2_x \right ) c_y + 2 s_x c_x s_y s_z e^{i\delta}}
~ e^{2i\delta} \cr
\frac{m_2}{m_3} e^{2i\sigma} =
+ \frac{1}{c_y c^2_z} \cdot \frac{c^2_x s^2_y \left (c^2_z - s^2_z \right )
- s_x c_y s^2_z \left (s_x c_y + 2 c_x s_y s_z e^{i\delta} \right ) e^{-2i\delta}}
{\left (s^2_x - c^2_x \right ) c_y + 2 s_x c_x s_y s_z e^{i\delta}}
~ e^{2i\delta} }$
\\ \\
$\rm E_2:
\left ( \matrix{
{\bf 0} & \times & \times \cr
\times & \times & \times \cr
\times & \times & {\bf 0} \cr} \right )$
&
$\matrix{
\frac{m_1}{m_3} e^{2i\rho} =
- \frac{1}{s_y c^2_z} \cdot \frac{s^2_x c^2_y \left (c^2_z - s^2_z \right )
- c_x s_y s^2_z \left (c_x s_y + 2 s_x c_y s_z e^{i\delta} \right ) e^{-2i\delta}}
{\left (s^2_x - c^2_x \right ) s_y - 2 s_x c_x c_y s_z e^{i\delta}}
~ e^{2i\delta} \cr
\frac{m_2}{m_3} e^{2i\sigma} =
+ \frac{1}{s_y c^2_z} \cdot \frac{c^2_x c^2_y \left (c^2_z - s^2_z \right )
- s_x s_y s^2_z \left (s_x s_y - 2 c_x c_y s_z e^{i\delta} \right ) e^{-2i\delta}}
{\left (s^2_x - c^2_x \right ) s_y - 2 s_x c_x c_y s_z e^{i\delta}}
~ e^{2i\delta} }$
\\ \\
$\rm E_3:
\left ( \matrix{
{\bf 0} & \times & \times \cr
\times & \times & {\bf 0} \cr
\times & {\bf 0} & \times \cr} \right )$
&
$\matrix{
\frac{m_1}{m_3} e^{2i\rho} =
- \frac{1}{c^2_z} \cdot \frac{s^2_x s_y c_y \left (c^2_z - s^2_z \right )
+ c_x s^2_z \left [c_x s_y c_y - s_x \left (s^2_y - c^2_y \right ) s_z e^{i\delta}
\right ] e^{-2i\delta}}
{\left (c^2_x - s^2_x \right ) s_y c_y + s_x c_x \left (c^2_y - s^2_y \right )
s_z e^{i\delta}} ~ e^{2i\delta} \cr
\frac{m_2}{m_3} e^{2i\sigma} =
+ \frac{1}{c^2_z} \cdot \frac{c^2_x s_y c_y \left (c^2_z - s^2_z \right )
+ s_x s^2_z \left [s_x s_y c_y + c_x \left (s^2_y - c^2_y \right ) s_z e^{i\delta}
\right ] e^{-2i\delta}}
{\left (c^2_x - s^2_x \right ) s_y c_y + s_x c_x \left (c^2_y - s^2_y \right )
s_z e^{i\delta}} ~ e^{2i\delta} }$
\\ \\
$\rm F_1:
\left ( \matrix{
\times & {\bf 0} & {\bf 0} \cr
{\bf 0} & \times & \times \cr
{\bf 0} & \times & \times \cr} \right )$
&
$\matrix{
\frac{m_1}{m_3} e^{2i\rho} = 1 \cr
\frac{m_2}{m_3} e^{2i\sigma} = 1 }$
\\ \\
$\rm F_2:
\left ( \matrix{
\times & {\bf 0} & \times \cr
{\bf 0} & \times & {\bf 0} \cr
\times & {\bf 0} & \times \cr} \right )$
&
$\matrix{
\frac{m_1}{m_3} e^{2i\rho} =
\frac{s_x c_y + c_x s_y s_z e^{-i\delta}}
{s_x c_y + c_x s_y s_z e^{+i\delta}} ~ e^{2i\delta} \cr
\frac{m_2}{m_3} e^{2i\sigma} =
\frac{c_x c_y - s_x s_y s_z e^{-i\delta}}
{c_x c_y - s_x s_y s_z e^{+i\delta}} ~ e^{2i\delta} }$
\\ \\
$\rm F_3:
\left ( \matrix{
\times & \times & {\bf 0} \cr
\times & \times & {\bf 0} \cr
{\bf 0} & {\bf 0} & \times \cr} \right )$
&
$\matrix{
\frac{m_1}{m_3} e^{2i\rho} =
\frac{s_x s_y - c_x c_y s_z e^{-i\delta}}
{s_x s_y - c_x c_y s_z e^{+i\delta}} ~ e^{2i\delta} \cr
\frac{m_2}{m_3} e^{2i\sigma} =
\frac{c_x s_y + s_x c_y s_z e^{-i\delta}}
{c_x s_y + s_x c_y s_z e^{+i\delta}} ~ e^{2i\delta} }$ 
\\ \\ 
\end{tabular}
\end{center}
\end{table}

\newpage

\begin{figure}[t]
\vspace{0cm}
\epsfig{file=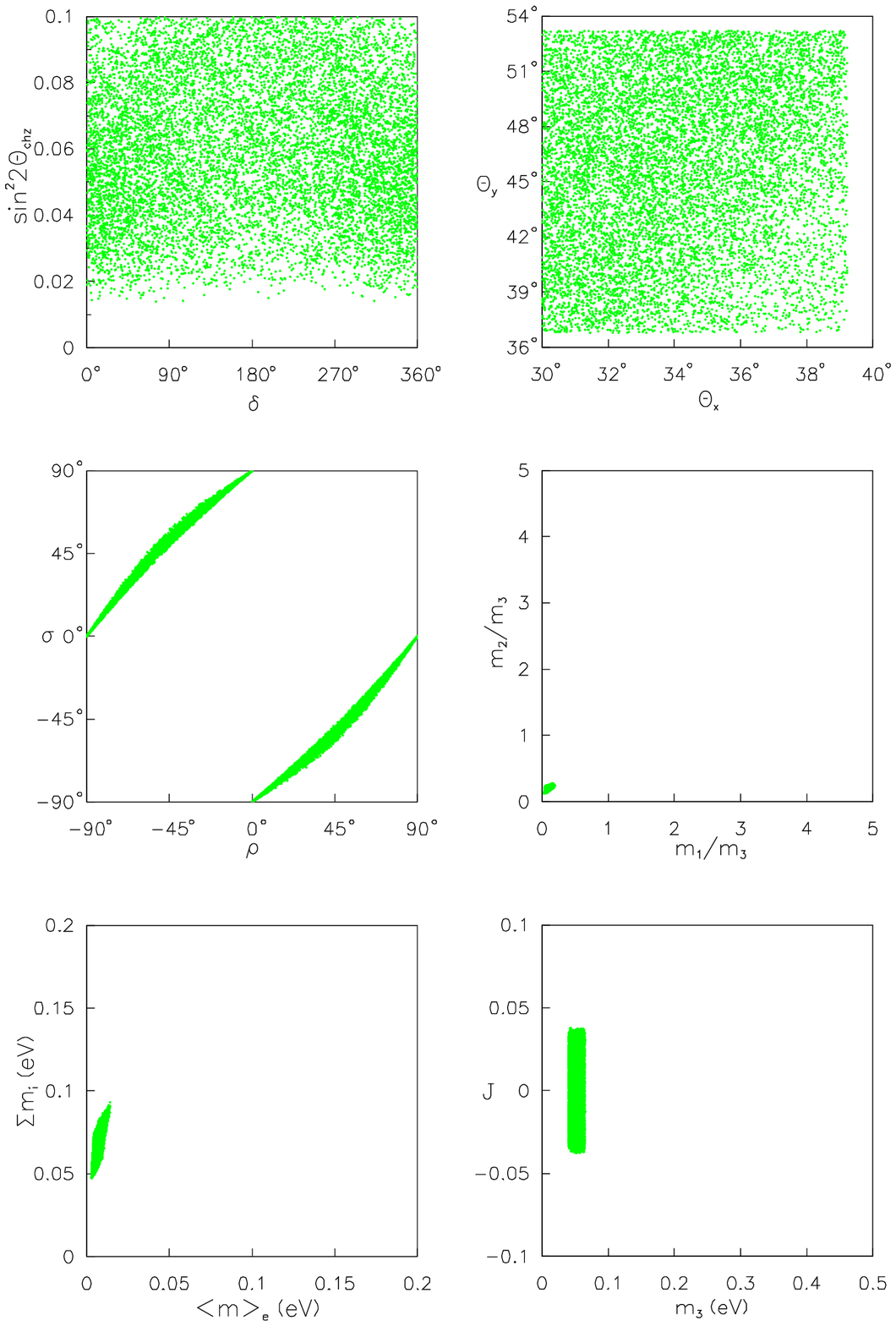,bbllx=1cm,bblly=2cm,bburx=17.5cm,bbury=28cm,%
width=14.5cm,height=21cm,angle=0,clip=0}
\vspace{-3cm}
\caption{Pattern $\rm A_1$ of the neutrino mass matrix $M$: allowed regions
of $\sin^2 2\theta_{\rm chz}$ versus $\delta$, $\theta_y$ versus $\theta_x$,
$\sigma$ versus $\rho$, $m_2/m_3$ versus $m_1/m_3$,
$\sum m_i$ versus $\langle m \rangle_e$, and $J$ versus $m_3$.}
\end{figure}

\begin{figure}[t]
\vspace{0cm}
\epsfig{file=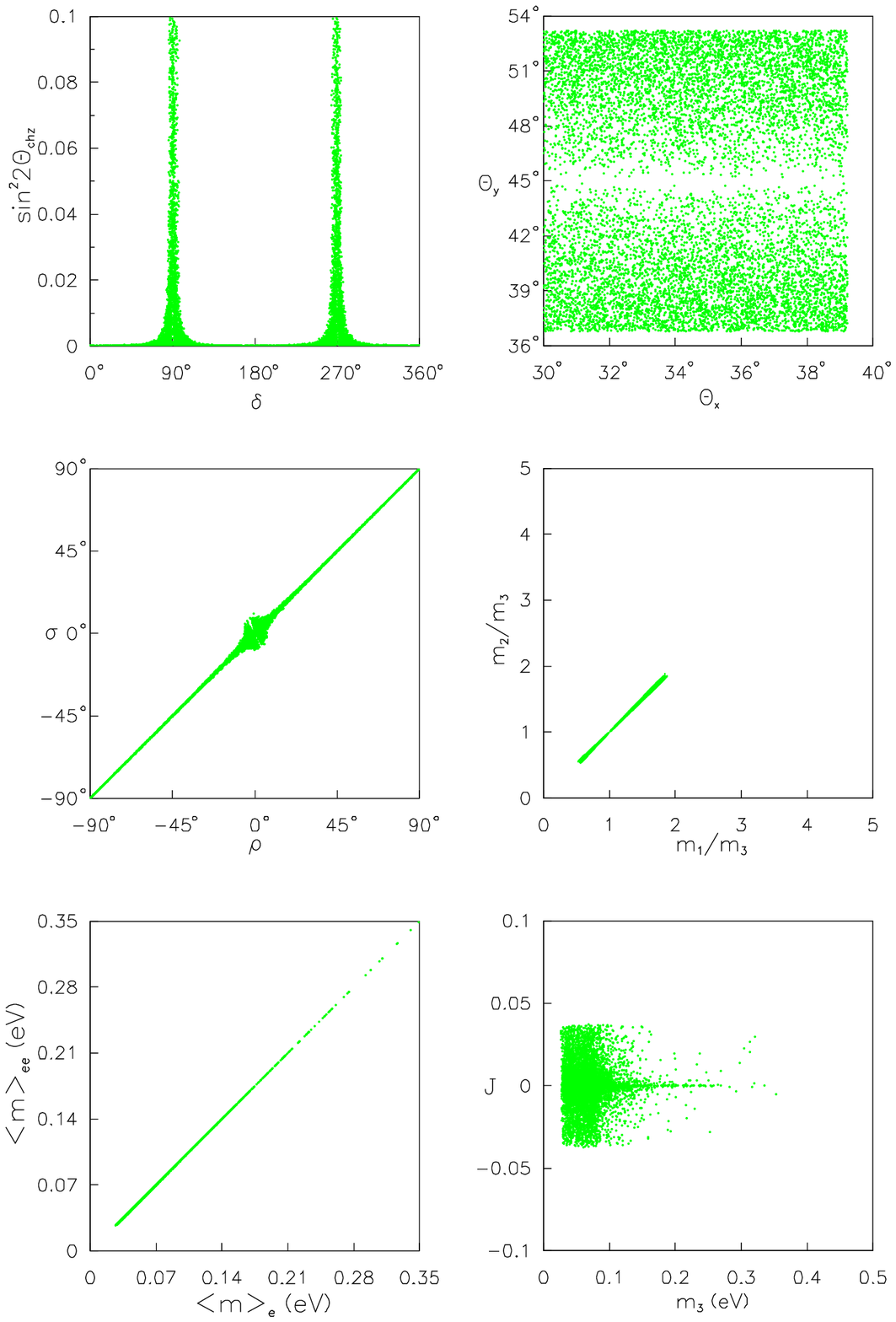,bbllx=1cm,bblly=2cm,bburx=17.5cm,bbury=28cm,%
width=14.5cm,height=21cm,angle=0,clip=0}
\vspace{-3cm}
\caption{Pattern $\rm B_1$ of the neutrino mass matrix $M$: allowed regions
of $\sin^2 2\theta_{\rm chz}$ versus $\delta$, $\theta_y$ versus $\theta_x$,
$\sigma$ versus $\rho$, $m_2/m_3$ versus $m_1/m_3$,
$\langle m \rangle_{ee}$ versus $\langle m \rangle_e$, and $J$
versus $m_3$.}
\end{figure}

\begin{figure}[t]
\vspace{0cm}
\epsfig{file=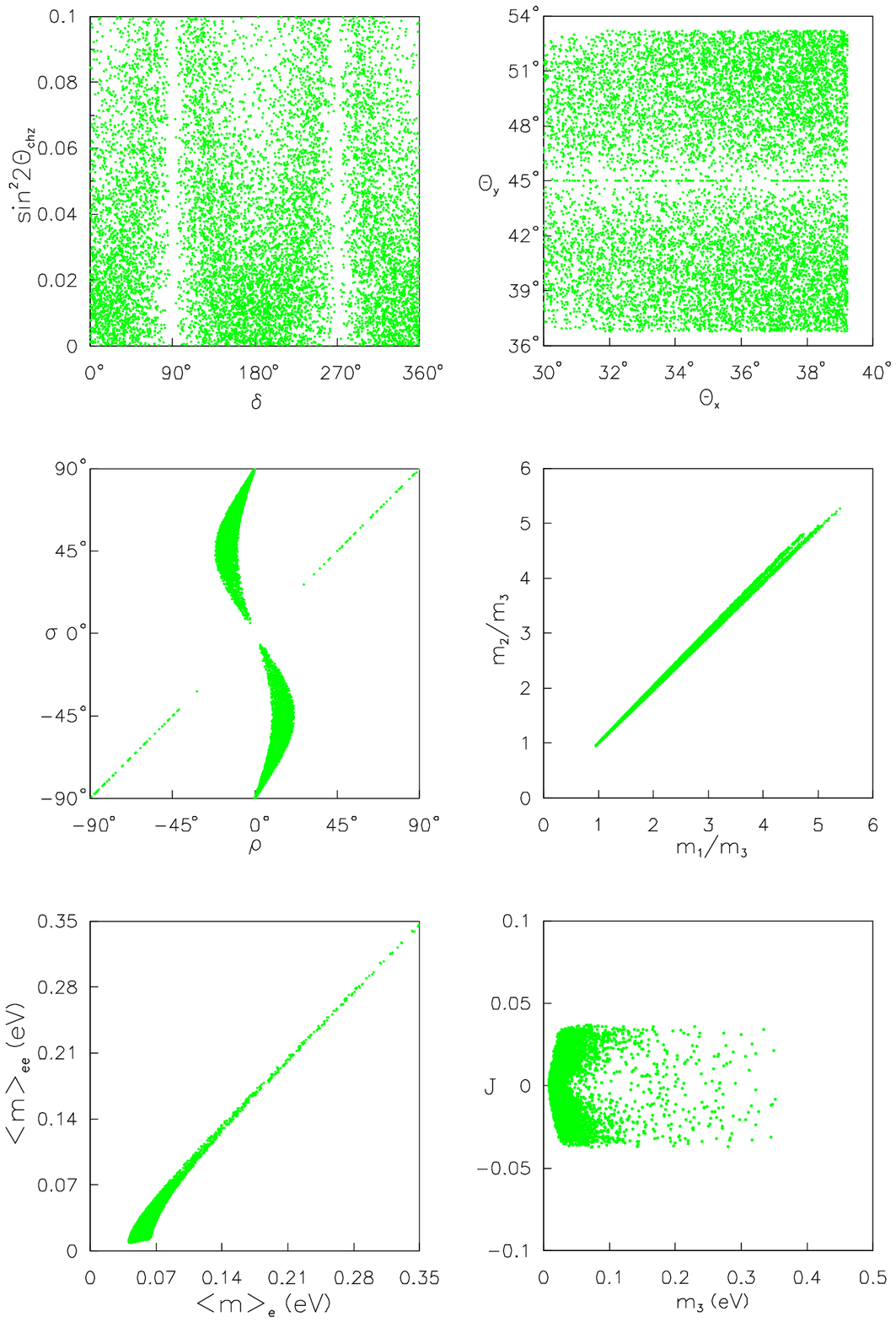,bbllx=1cm,bblly=2cm,bburx=17.5cm,bbury=28cm,%
width=14.5cm,height=21cm,angle=0,clip=0}
\vspace{-3cm}
\caption{Pattern $\rm C$ of the neutrino mass matrix $M$: allowed regions
of $\sin^2 2\theta_{\rm chz}$ versus $\delta$, $\theta_y$ versus $\theta_x$,
$\sigma$ versus $\rho$, $m_2/m_3$ versus $m_1/m_3$,
$\langle m \rangle_{ee}$ versus $\langle m \rangle_e$, and $J$
versus $m_3$.}
\end{figure}

\begin{figure}[t]
\vspace{0cm}
\epsfig{file=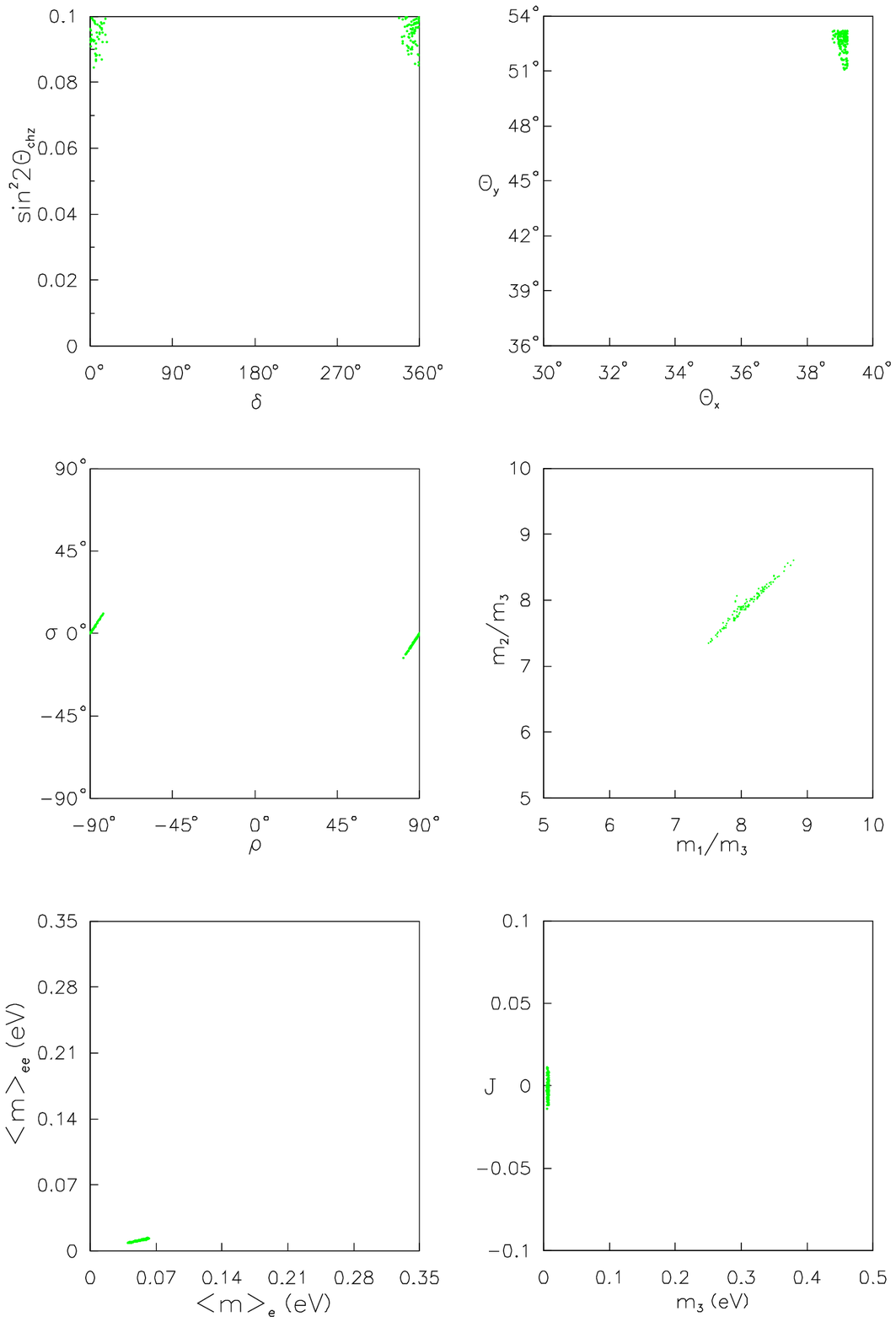,bbllx=1cm,bblly=2cm,bburx=17.5cm,bbury=28cm,%
width=14.5cm,height=21cm,angle=0,clip=0} \vspace{-3cm}
\caption{Pattern $\rm D_1$ of the neutrino mass matrix $M$:
allowed regions of $\sin^2 2\theta_{\rm chz}$ versus $\delta$,
$\theta_y$ versus $\theta_x$, $\sigma$ versus $\rho$, $m_2/m_3$
versus $m_1/m_3$, $\langle m \rangle_{ee}$ versus $\langle m
\rangle_e$, and $J$ versus $m_3$.}
\end{figure}

\end{document}